\documentclass[pra,superscriptaddress,amsmath,amssymb,twocolumn]{revtex4-1}
\pdfoutput=1

\usepackage{amsthm}
\usepackage{graphicx}
\usepackage[all]{xy}

\newcommand{\mc}[1]{\mathcal #1}

\newcommand{\ave}[1]{\langle #1 \rangle}

\newcommand{\tr}{{\rm Tr}\,}
\newcommand{\one}{{\bf 1}}

\newcommand{\mysection}[1]{\section{#1}}

\newcommand{\chan}[1]{\mathcal #1}

\newcommand{\op}[1]{#1}

\newcommand{\manifold}{{\mathcal K}}
\newcommand{\cfun}{f}
\newcommand{\unemph}[1]{#1}

\theoremstyle{plain}

\theoremstyle{definition}

\begin{document}

\title{Information geometric approach to the renormalisation group}

\author{C\'edric B\'eny}
\affiliation{Institut f\"ur Theoretische Physik, Leibniz Universit\"at Hannover, Appelstr. 2, 30167 Hannover, Germany}
\author{Tobias J.\ Osborne}
\affiliation{Institut f\"ur Theoretische Physik, Leibniz Universit\"at Hannover, Appelstr. 2, 30167 Hannover, Germany}
\affiliation{Riemann Center for Geometry and Physics, Leibniz Universit\"at Hannover, Appelstr. 2, 30167 Hannover, Germany}

\begin{abstract}
We propose a general formulation of the renormalisation group as a family of quantum channels which connect the microscopic physical world to the observable world at some scale. By endowing the set of quantum states with an operationally motivated information geometry, we induce the space of Hamiltonians with a corresponding metric geometry. The resulting structure allows one to quantify information loss along RG flows in terms of the distinguishability of thermal states. In particular, we introduce a family of functions, expressible in terms of two-point correlation functions, which are non increasing along the flow. Among those, we study the speed of the flow, and its generalization to infinite lattices.   
\end{abstract}

\maketitle

The renormalisation group (RG), one of the most profound ideas in science, allows us to understand long-range physics without requiring us to completely describe the fundamental constituents of the universe. This is because small-scale microscopic behaviour can be absorbed by a handful of parameters of some effective theory.  

The RG is typically taken to act on the space of \unemph{theories} or, more concretely, the space $\manifold$ of \emph{Hamiltonians}, and is thus a kind of \unemph{superoperation} which produces from a given initial Hamiltonian an effective Hamiltonian at some larger scale. By repeatedly composing the RG operation one generates a \unemph{flow} or \unemph{trajectory} in $\manifold$. This flow typically possesses attractive fixed points/surfaces corresponding to the physics at large lengthcales. 

The fact that vastly different theories can converge to the same submanifold characterized by a handful of relevant parameters --- the phenomenon of ``\unemph{universality}'' --- invites an information-theoretic explanation~\cite{preskill00}. Indeed, because the renormalisation procedure \emph{ignores} progressively larger scale features of the theory, its irreversibility should be directly related to a loss of information about the microscopic physics. 

Previous work in this direction has focused on the attempt to derive a function on the manifold $\manifold$ which always decreases along the RG flow, except at a fixed point, which would prove the irreversibility of the renormalisation procedure.
For $1+1$-dimensional quantum fields, this is achieved by \unemph{Zamolodchikov's $c$-theorem}~\cite{zamolodchikov86}, which establishes the existence of such a scalar function, expressible in terms of the correlation functions of the theory. 
Similar results have been obtained for higher dimensional theories~\cite{cardy88,komargodski11}. More direct information-theoretic approaches were also explored in the context of classical field theory~\cite{gaite95,apenko12}, or using ground-state entanglement~\cite{casini07}.

Here we propose a different approach based on an information-theoretic formulation of the renormalisation group.
We argue that the RG is naturally expressed in terms of a markov process on the manifold of statistical quantum states. Because thermal states essentially specify their Hamiltonian, we can pull back any information-theoretic structure on the manifold of statistical quantum states to the manifold of Hamiltonians. In this context, the information being lost along the RG flow is that which measures \emph{distinguishability} between different theories. This gives rise to an information metric on $\manifold$ which allows for the derivation of QFT-friendly $c$-like quantities. 

\begin{figure}
\includegraphics[width=0.75\columnwidth]{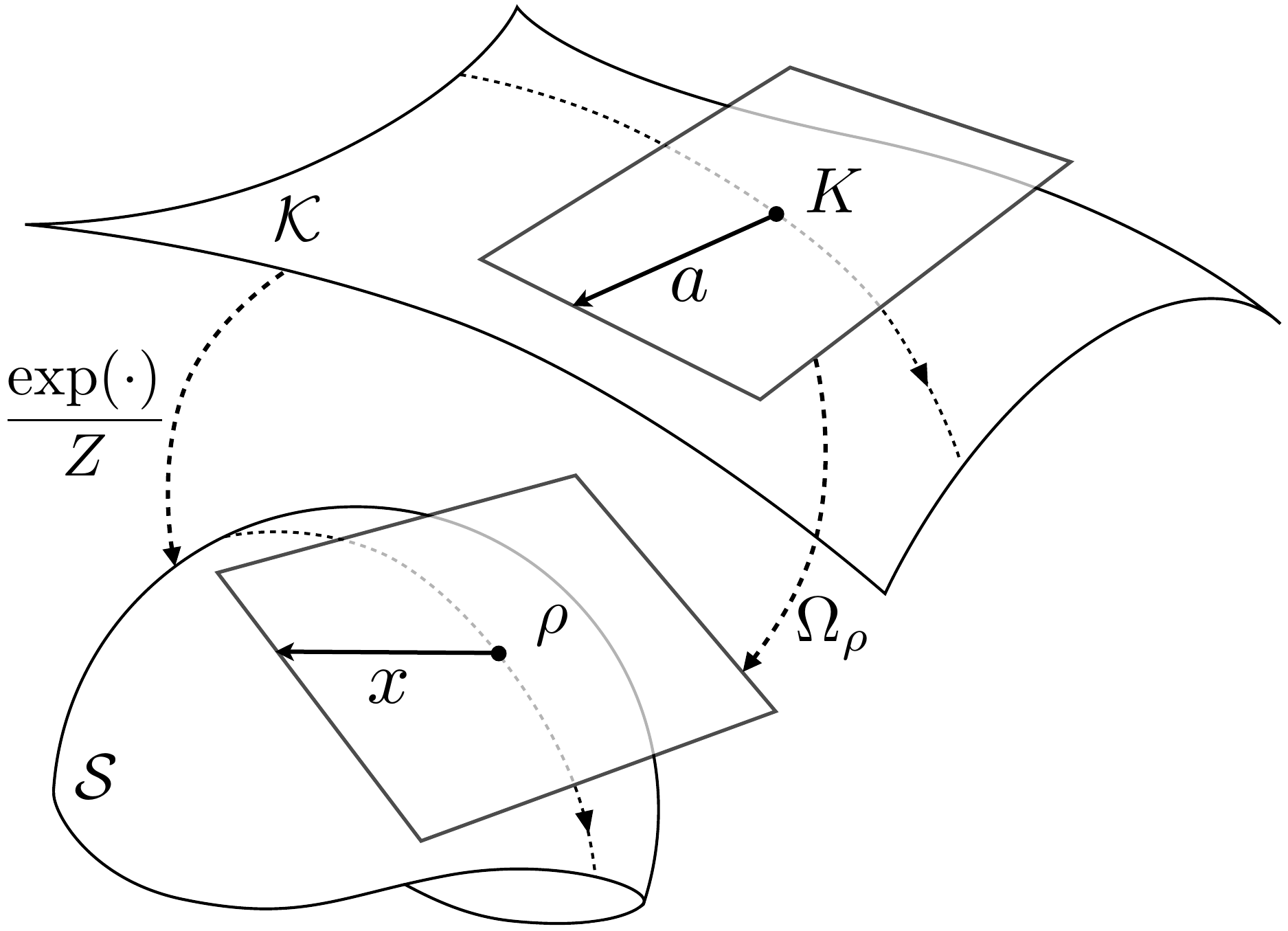} 
\caption{Artist's impression of the manifold of states $\mathcal S$ and the manifold of Hamiltonians $\manifold$. The manifolds are related by the exponential map, and their tangent spaces by its derivative $\Omega_\rho$. Here $\rho = e^K/Z$ and $x = \Omega_\rho(a)$.}
\label{diagram}
\end{figure}

Suppose that the fundamental constituents of the system under study are quantum particles/degrees of freedom interacting at some extremely short length scale $\varepsilon$. We write the (possibly mixed) quantum state of this system as $\rho_{\varepsilon}$. When we perform an experiment at a larger length scale $\ell \gg \varepsilon$ we are making a very imperfect measurement: we are ignoring many unobservable degrees of freedom. 
The most general procedure to ``ignore'' degrees of freedom consistent with the laws of quantum mechanics is described by a \emph{quantum channel} $\mathcal{E}$, i.e., a completely positive and trace-preserving map. We thus write the best description of the quantum state available to our experiments as $\rho_{\ell} = \mathcal{E}_\ell(\rho_{\varepsilon})$, where $\mathcal{E}_\ell$ denotes the quantum channel which ignores (or discards) all the degrees of freedom we cannot observe. Typically in the context of Wilsonian renormalisation~\cite{wilson75} and the functional RG the map $\mathcal{E}_{\ell}$ is the partial trace channel over all momenta $|k| \gg 1/\ell$. For our purposes, we suppose that knowing the effective state at any intermediate scale is enough to obtain the effective state at a larger scale. 
We call $\ell \mapsto \mathcal{E}_\ell$ the \emph{renormalisation ``group''} or \emph{process}.

In order to connect this flow on states to a flow on Hamiltonians, we assume that the fundamental state $\rho_\varepsilon$ is in thermal equilibrium at a temperature $\tau$, and hence has the form $\rho_\varepsilon = e^{-H_\varepsilon/\tau}/Z$ for some Hamiltonian $H_\varepsilon$. Furthermore, if the temperature is defined by a reservoir with which our system is in contact, it is natural that the effective states $\rho_\ell=\chan E_\ell(\rho_\varepsilon)$ are thermal states at the same temperature $\tau$. The effective Hamiltonian $H_\ell$ at scale $\ell$ can then be defined so that $\rho_\ell = e^{H_\ell/\tau}/Z$, or $H_\ell = - \tau \log (Z \rho_\ell)$. The partition function is independent of $\ell$, thanks to the fact that the maps $\chan E_\ell$ are trace preserving. 
Since the temperature is assumed to be constant, we can work with the operator $K_\ell = -H_\ell/\tau$ instead of $H_\ell$ and, abusing language, call those {\it Hamiltonians}. 
The basic arena of our study is therefore the manifold $\manifold$ of Hermitian operators $K$ with the same $Z = \tr e^K$.

On a finite-dimensional Hilbert space, the process $\mathcal E_\ell$ is generated by a Lindblad superoperator $\mc L_\ell$, so that  
\(
\frac{d}{d\ell} \rho_\ell = \mc L_\ell(\rho_\ell).
\)
In order to further simplify the presentation, we assume a constant generator $\mathcal L$. This may require the parameter of the flow to be different from the lengthscale $\ell$, albeit still a monotone function of $\ell$. In order to differentiate between them, we therefore denote that parameter $t$, and write
\(
\frac{d}{dt} \rho_t = \mc L(\rho_t),
\)
so that the family of channels $\chan E_t = e^{t \mathcal L}$ is the semigroup generated by $\mathcal L$.

We define the effective Hamiltonian at the scale indexed by $t$ to be $K_t$ such that $\rho_t = e^{K_t}/Z$. The evolution of the Hamiltonian $K_t = \log (Z \rho_t)$ under the renormalisation group is given by
\[
\frac{d}{dt} K_t = \mc B(K_t)
\]
with tangent field
\(
\mc B(K) = \Omega_{e^{K}}^{-1} \mc L(e^{K}) = \Omega_{\rho}^{-1} \mc L(\rho),
\)
which we expressed in terms of the useful linear superoperator~\cite{ruskai05}
\begin{equation}
\begin{split}
\Omega_{\rho}^{-1}(\op x) &= \frac{d}{dt} \log(\rho + t \op x)\bigr|_{0}
= \int_0^\infty du \frac 1 {\rho + u \one} \op x  \frac 1 {\rho + u \one}.
\end{split}
\end{equation}
This superoperator can be understood as a noncommutative generalization of the operation of dividing by $\rho$. 
In geometric terms, it pushes forward a tangent vector at $\rho = e^K/Z$ to the corresponding tangent vector at $K$.  
We also need its inverse
\[
\begin{split}
\Omega_{\rho}(\op a) = \frac{d}{dt} e^{\log \rho + t \op a}\bigr|_{t=0} = \int_0^1 \rho^{s} \op a \,\rho^{(1-s)} ds.
\end{split}
\]

On the manifold of states, since the diffeomorphism $e^{t \mc L}$ is linear, its push-forward on tangent vectors is just itself. Hence, on the manifold of Hamiltonians, it maps the tangent vector $\op a$ at $K$ to
\(
\op a_t = \Omega_{\rho_t}^{-1} e^{t \mc L}\, \Omega_{\rho} (\op a),
\)
at the point $K_t$. 
The infinitesimal form is
\begin{equation}
\label{push2}
\begin{split}
\frac{d}{dt}\op a_t &= \frac{d}{d\epsilon} \mc B(K_t + \epsilon \, \op a_t)\bigr|_{\epsilon=0}\\
&= \left[{ \Omega_{\rho_t}^{-1} \mc L \,\Omega_{\rho_t} - \Omega_{\rho_t}^{-1} \dot \Omega_{\rho_t}}\right] (\op a_t)\\
&=: \nabla_{\rho_t} a_t,
\end{split}
\end{equation}

Although our set-up is finite-dimensional, it would be natural to identify $\mc B(K)$ with the \emph{beta-function} of quantum field theory in the following way.
For a translation-invariant QFT one would usually assume that $K$ has the form 
$K = \int dx \sum_i g_i \Phi_i(x)$, where the family $\{\Phi_i(x)\}_{i=0}^\infty$ spans all local field operators at $x$, and $g_i$ are the corresponding coupling constants. The generator $\mc B(K)$ then should have the form
\(
\mc B(K) =  \int dx \, \sum_i \beta_i(K)\, \Phi_i(x),
\)
where the component $\beta_i(K) \in \mathbb R$ is the beta-function for the coupling constant $g_i$. 

\mysection{Information geometry}
Recall that vectors tangent to the Hamiltonian $K$ are observables with zero expectation value with respect to the state $e^K/Z$. For any two such observables $\op a$ and $\op b$, we define the bilinear form  
\begin{equation}
\label{metric}
\begin{split}
\ave{\op a, \op b}_K &:= -\frac{\partial^2}{\partial t\, \partial s}F(K + t \op a + s\op b)\Bigr|_{0}
= \tr(\op a \, \Omega_\rho( \op b)),
\end{split}
\end{equation}
where 
\(
F(K) = - \log \tr e^K
\)
is the free energy for the Hamiltonian $H = - \tau K$. 

This defines a Riemannian metric on the manifold of observables. The corresponding metric on the manifold of states is the Kubo-Mori metric: Given two tangent (traceless) vectors $\op x$ and $\op y$ at $\rho$,
\begin{equation}
\label{statemetric}
\begin{split}
\ave{\op x, \op y}_\rho &= \frac{\partial^2}{\partial t\, \partial s} S(\rho + t \op x + s \op y \| \rho )\Bigr|_{0}
=  \tr(\op x \, \Omega_\rho^{-1}( \op y)),
\end{split}
\end{equation}
where
\(
S(\rho' \| \rho) = \tr[\rho' (\log \rho' - \log \rho) ]
\)
is the {\em relative entropy}~\footnote{This duality between the free energy and the relative entropy can be traced to the fact that they are Legendre transforms of each other~\cite{hiai93}.}, an information-theoretic quantity which measures how {\em distinguishable} the two states $\rho$ and $\rho'$ are~\cite{hiai91}. 
This gives an operational meaning to our metric. We can think of the norm $\sqrt{\ave{\op a,\op a}}_K$ of a small tangent vector $\op a$ as measuring how distinguishable the perturbed state $e^{K + \op a}/Z$ is from $e^{K}/Z$.

There are two compatible interpretations of the relation between these bilinear forms, connected to the dual role played by observables in equilibrium statistical mechanics. On one hand, observables are perturbations of Hamiltonians, and are hence tangent vectors to the manifold of Hamiltonians, connected to the manifold of states via the exponential diffeomorphism. On the other hand, observables can be seen as cotangent vectors on the manifold of states, since they are meant to be combined with states to yield an expectation value. 

The main property of these dual metrics is that the norm of a tangent vector cannot increase under the action of a quantum channel~\cite{petz94}. 
On states this means that for any channel $\chan N$ and tangent vector $\op x$,
\begin{equation}
\ave{\chan N(\op x), \chan N(\op x)}_{\chan N(\rho)} \le \ave{\op x, \op x}_{\rho}.
\end{equation}
In classical probability theory, there exists only one metric with this property: the Fisher-Rao metric, which has been extensively used in statistical physics (e.g. see Refs.~\cite{diosi84,brody97,dolan98} in the context of renormalization). The metric defined in Eq.~\eqref{statemetric} is one of a family of noncommutative generalizations~\cite{petz96}, and reduces to the Fisher-Rao metric for commuting operators and states.

It follows from this contraction property that, along the flow generated by $\mc B$ on the manifold of Hamiltonians, for any tangent vector $\op a$,
\begin{equation}
\label{contract0}
\frac{d}{dt} \ave{\op a_t,\op a_t}_{K_t} \le 0,
\end{equation}
which, more explicitly, is equivalent to the statement that for all $\rho = e^K/Z$, and for all tangent vectors $\op a$,
\begin{equation}
\label{contract}
2 \,\ave{ \op a,\nabla_\rho (\op a)}_K + \tr(\op a\, \dot \Omega_\rho (\op a)) \le 0,
\end{equation}
where the $\dot \Omega_{\rho}$ denotes the derivative of $\Omega_{\rho}$ with respect to $t$.

\mysection{Non-increasing scalar}
Motivated by Zamolodchikov's $c$-theorem~\cite{zamolodchikov86}, we want to build a scalar field $K \mapsto \cfun(K)$ on the manifold $\manifold$ of Hamiltonians which has the property that it never increases under the renormalisation flow generated by $\mc B$, i.e. such that
\begin{equation}
\label{mono}
\frac{d}{dt} \cfun(K_t) \le 0
\end{equation}
and, if possible, such that it stops decreasing only at a fixed point. 

In the case where an exact fixed point of the renormalisation semigroup is known, i.e., a state $\sigma$ such that $\mathcal L(\sigma) = 0$, then any contractive measure of distinguishability with respect to $\sigma$ has that property. An example would be the relative entropy: $\cfun(K) = S(e^{K}/Z \| \sigma)$. Indeed, $S(\rho_1\|\rho_2)$ satisfies $S(\chan E(\rho_1)\|\chan E(\rho_2)) \le S(\rho_1\|\rho_2)$ for any quantum channel $\chan E$. It has a straightforward interpretation: it measures the distinguishability with respect to the fixed point. Moreover, if $\sigma$ is the maximally mixed state, i.e. the infinite temperature fixed point, then this function is simply a constant minus the entropy. However, this presupposes that the fixed point towards which the system flows is known. In addition, it is a very difficult quantity to compute. We propose instead a quantity which can be computed locally directly in terms of correlation functions.

In view of Eq.~\eqref{contract0}, the norm of any observable, which is expressed in terms of a Green function, does decrease along the flow. That quantity is not purely a function of the Hamiltonian $K$. However, it can be made so if the observable is replaced by a tangent \emph{field} $K \mapsto A(K)$ which is invariant under the flow, i.e. such that, for all $t$, $A(K_t)$ is equal to the image under the flow of the tangent vector $A(K_0)$. Using Eq.~\eqref{push2}, the infinitesimal version of this condition is
\begin{equation}
\label{inv}
\frac{d}{dt}A(K_t) = \nabla_{\rho_t}(A(K_t)),
\end{equation}
where $\rho_t = e^{K_t}/Z$. 
This is equivalent to the statement that the Lie derivative of $A$ with respect to $\mc B$ is zero. 
An example is $A=\mc B$ itself, or fields of the form $A(K) = \Omega_{e^K}^{-1}{\mc L'}(e^K)$, where $\mc L'$ is any linear superoperator which commutes with $\mc L$. 
Indeed, if a field $A$ satisfies Eq.~\eqref{inv}, then Eq.~\eqref{contract} implies that the scalar 
\begin{equation}
\label{cquant}
\cfun(K) = \ave{A(K),A(K)}_K
\end{equation}
where $\rho = e^K/Z$, satisfies Eq.~\eqref{mono}. For instance, if we use $A:=\mc B$, then $\cfun(K)$ is the squared {\em speed} of the flow, 
\(
\cfun(K) = \ave{\mc B(K),\mc B(K)}_K
\)
measured in the distinguishability metric, which is zero if and only if $\mc B(K) = 0$, i.e. at a fixed point. In the classical setting, this corresponds to the ``velocity function'' of Ref.~\cite{brody97}.

\mysection{Infinite lattices}
We have so far worked within the framework of finite-dimensional Hilbert spaces, and for a generic semigroup $\mc E_t$. However, to study renormalisation more specifically, we may need to consider infinite lattices and translation-invariant states, for which quantities like those defined in Eq.~\eqref{cquant} are usually infinite.

The natural course of action is to consider a {\em density} corresponding to $\cfun(K)$. But, as we will see, this presents severe issues if the renormalisation flow includes an explicit rescaling of the lattice.  

Therefore, we first consider the case where the renormalisation flow commutes with translations, which means that it consists only of decimation with no active rescaling of length. Explicitly, if $T_x$ is the superoperator implementing a translation of the state by the lattice vector $x$, we require that
\(
\mc L\, T_x = T_x \mc L.
\)

In addition, we require that the renormalisation flow stays inside a submanifold $\mc M$ of local translation invariant Hamiltonians, which implies that, for any $K \in \mc M$, 
\(
\mc B(K) = \sum_x \op b_x(K),
\)
where the operators $\op b_x(K) = T_x(\op b(K))$ act only on a finite number of sites, and $T_x$ is a linear superoperator implementing a translation by the lattice vector $x$.

We then define
\begin{equation}
\label{betadensity}
\begin{split}
\cfun(K) &= \sum_{x} \ave{\op b_0(K), \op b_x(K)}_K = \ave{\op b(K), \mc B(K)}_K.\\
\end{split}
\end{equation}
This quantity corresponds to the speed {\em density} of the flow. We show that under the above assumptions, it is indeed non-increasing along the flow, i.e.
\(
\frac{d}{dt} \cfun(K_t) \le 0.
\)
First we observe that, since $[T_x,\nabla_\rho] =0$ and $T_x(\rho)=\rho$, then
\begin{equation}
\label{aux1}
\ave{\nabla_\rho \op b_y, \op b_x} = \ave{\nabla_\rho T_y b, \op b_x} = \ave{\nabla_\rho b, T_{-y} \op b_x}= \ave{\nabla_\rho b, \op b_{x-y}}.
\end{equation}
Let 
\(
\mc B^L := \sum_{|x|<L} \op b_x(K),
\)
and $V_L$ be the number of translations satisfying $|x|<L$. Then Eq.~\eqref{contract} implies
\[
\begin{split}
0 &\ge 2 \ave{\nabla_\rho \mc B^L, \mc B^L } + \tr(\mc B^L \dot \Omega_\rho(\mc B^L)) \\
&= \sum_{|y| \le L , |x|\le L}2\ave{\nabla_\rho \op b,\op b_{x-y}} + \tr(\op b \,\dot \Omega_\rho(\op b_{x-y}))\\
&= \sum_{x} f_L(x) \left[{ 2\ave{\nabla_\rho b,\op b_{x}} + \tr(\op b \,\dot \Omega_\rho(\op b_{x})) }\right],\\
\end{split}
\]
where $f_L(x)$ counts the number of identical elements in the double sum. It is such that
\[
\begin{split}
\lim_{L \rightarrow \infty} &\frac{1}{V_L}\sum_{x} f_L(x) \left[{ 2\ave{\nabla_\rho b,\op b_{x}} + \tr(\op b \,\dot \Omega_\rho(\op b_{x})) }\right]\\
&= 2\ave{\nabla_\rho b,\mc B} + \tr(\op b\, \dot \Omega_\rho(\mc B))
\end{split}
\]
which then must also be negative. We conclude by showing that $\frac{d}{dt} \cfun(K_t)$ is equal to that last quantity. 
The derivative $\frac{d}{dt} \cfun(K_t)|_{t=0}$ is made of three terms, one of which is equal to $\tr(\op b\, \dot \Omega_\rho(\mc B))$. The second term is just $\ave{b,\frac{d}{dt}\mc B|_{0}}= \ave{b,\nabla\mc B} = \ave{\nabla b,\mc B}$ by virtue of Eq.~\eqref{aux1}. The last term is  
\[
\begin{split}
\ave{\frac{d}{dt}\op b(K_t)\bigr|_0, \mc B} &= \sum_x \ave{\frac{d}{dt}\op b(K_t)\bigr|_0, T_x \op b} 
= \ave{\frac{d}{dt}\mc B(K_t)\bigr|_0, \op b}\\
&= \ave{\nabla_\rho(\mc B), \op b} = \ave{\nabla_\rho(\op b), \mc B},
\end{split}
\]
which concludes the proof.

\mysection{Scaling}

The quantity defined in Eq.~\eqref{betadensity} may fail to be non-increasing if the flow does not commute with translations. 
A simple counterexample is given by the one-dimensional Ising model with the flow generated by the superoperator
\(
\mc L(\rho) = \sum_i \left[{ \tr_i(\rho) - \rho }\right],
\)
where the sum is over all lattice sites and $\tr_i$ traces out the $i$th site and then, for all $j\ge i$, replaces the $j$th sites by the states of the $(j+1)$th site, i.e.\ it shifts the spins on the right of the $i$th in order to fill the gap. 
A good property of this renormalisation group is that it does not increase the range of the interactions. Hence the nearest-neighbour Ising Hamiltonian stays within the manifold of commuting translation-invariant Hamiltonians acting only on nearest-neighbours, which is three-dimensional. The Hamiltonians can be written as $K(J,h,c) = J \sum_i S_i S_{i+1} + h \sum_i S_i + c \one$, where $S_i$ is the diagonal matrix acting on site $i$ with eigenvalues $+1$ and $-1$. 
In finite dimensions, the expectation value of $\mc B(K)$ would alway
s be zero provided that the semigroup is trace-preserving. However, here one needs to choose the parameter $c$ so that the free energy density is zero, i.e. $c = -\log Z$.

On the stable submanifold defined by $c=-\log(2\cosh J)$ and $h=0$, we have $\mc B(K) = \sum_i b_i(K)$ with
\(
\op b_i(K(J)) = - e^{-J} \sinh(J) S_i S_{i+1} - \log(2 \cosh(J)) \one
\)
~\footnote{Incidentally, this is the same flow as the one derived in Ref.~\cite{brody97} with a different approach.}.
We then obtain
\(
\cfun(K) = e^{-2J} \left[{ \tanh(J) }\right]^2.
\)
One can check $J>0$ always decreases along the flow, but $\cfun(K)$ attains a maximum for a finite $J$, which indeed contradicts the proposition that $\cfun(K)$ be always decreasing along the flow.

Generally, the problem with an active scaling is that the {\em causal past} $\Sigma'$ of a region $\Sigma$ (with respect to the ``dynamics'' $\mc E_\ell$)  would scale as the volume of $\Sigma$ rather than its area, and therefore the difference between $\frac 1 {\Sigma}(\rho_\ell)_\Sigma$ and $\frac{1}{\Sigma'}(\rho_\ell)_{\Sigma'}$ does not necessarily vanish for large $\Sigma$.  

This is a problem because we are generally interested in renormalisation groups which converge to interesting fixed points. If no scaling is involved together with the removal of local information, then, strictly speaking, the only possible fixed points contain no correlations. 

Let us sketch a possible way of addressing this problem. For simplicity, instead of the differential state-independent quantity $\cfun(K)$, we consider the relative entropy density 
\[ 
D(\rho \| \sigma) = \lim_{\Sigma} \frac{1}{|\Sigma|} S(\rho_\Sigma \| \sigma_\Sigma),
\]
 where the limit is over the net of subsets $\Sigma$ of lattice sites, and $\rho_\Sigma$ denotes the reduced state on $\Sigma$ for a full state $\rho$.  

The idea is that, instead of implementing the scaling explicitly in the dynamics $\mc E_\ell$, which is problematic as it requires an unbounded generator as in the previous example, we can more simply implement it in the way that we compare states at different values of $\ell$. Because we are dealing with a relative entropy {\em density}, it has a unit: that of inverse distance. This unit should scale with $\ell$ appropriately. Assuming that $\mc L_\ell$ is chosen so that $\mc E_\ell$ erases information about a {\em scale} $\ell$, so that if $\epsilon$ is the lattice spacing then $\mc E_\epsilon(\rho) = \rho$, we define
\[
D_\ell(\rho_\ell \| \sigma_\ell) := \ell^d D(\rho_\ell \| \sigma_\ell),
\] 
where $d$ is the number of spatial dimensions. This quantity has no unit.  
This tool allows us to say that two states $\rho$ and $\sigma$ converge ``to the same fixed point'' if $\lim_{s \rightarrow \infty} D_\ell(\rho_\ell\| \sigma_\ell) = 0$. The fixed point itself is represented only as the resulting equivalence class of states.

This quantity does not generally decrease with $\ell$. Indeed, we have
\begin{equation}
\label{someequ}
\frac{d}{d\ell} D_\ell(\rho_\ell \| \sigma_\ell) = \frac{d}{\ell} D_\ell(\rho_\ell \| \sigma_\ell) + \ell^d \frac{d}{d\ell} D(\rho_\ell \| \sigma_\ell).
\end{equation}
In general we only expect that the last term is negative (or zero), but we don't know that it is negative {\em enough} to compensate for the positivity of the first term $ \frac{d}{\ell} D_\ell(\rho_\ell \| \sigma_\ell) $.  However, a simple argument suggests that this term can be canceled by adding the right amount of depolarisation to the coarse-graining procedure. Let $\mc K_\ell(\rho) = \frac{d}{\ell} \sum_i (  \mc D_i(\rho) - \rho)$, where $\mc D_i$ is the application of the depolarisation map $\mc D(\rho) = \tr(\rho) \one /\tr \one$ to the $i$th site, and $d$ is the dimension of space.  Let $\rho_\ell$ is the state evolved under the dynamics generated by $\mc L_\ell' = \mc L_\ell + \mc K_\ell$. Furthermore, we write $\rho_{\ell,\epsilon} = \rho_\ell + \epsilon \mc K_\ell(\rho_\ell)$.
From the joint convexity of the relative entropy we obtain, to first order in $\epsilon$,
\[
\begin{split}
S((\rho_{\ell,\epsilon} )_\Sigma \| ( \sigma_{\ell,\epsilon})_\Sigma) \le &\left({1-\epsilon |\Sigma| \frac d \ell}\right) S((\rho_\ell)_\Sigma \|(\sigma_\ell)_\Sigma  ) \\
&+ \epsilon \frac d \ell \sum_i S((\rho_\ell)_{\Sigma \backslash i }\|(\sigma_\ell)_{\Sigma \backslash i })\\
&= S_\Sigma   - \epsilon \frac d \ell \sum_i \left({ S_{\Sigma} -  S_{\Sigma \backslash i}  }\right)\\
\end{split}
\]
where we abbreviated $S_\Sigma = S((\rho_\ell)_\Sigma \|(\sigma_\ell)_\Sigma  )$, and $\Sigma\backslash i$ is the region $\Sigma$ without the site $i$.
Since also the derivate of the relative entropy is negative with respect to the generator $\mc L_\ell$, we have altogether
\[
\frac{d}{d\ell} \frac{1}{|\Sigma|} S_\Sigma \le - \frac d \ell \frac{1}{|\Sigma|} \sum_i \left({ S_{\Sigma} -  S_{\Sigma \backslash i}  }\right).
\]
Since, for large $\Sigma$ we expect $S_\Sigma \simeq |\Sigma| D(\rho_\ell \| \sigma_\ell)$, then in the thermodynamic limit, we ought to have
\[
\frac{d}{d\ell} D(\rho_\ell \| \sigma_\ell) \le - \frac d s D(\rho_\ell \| \sigma_\ell),
\]
which, using Equ.~\ref{someequ}, would imply that
\(
\frac{d}{d\ell} D_\ell(\rho_\ell \| \sigma_\ell) \le 0.
\)

\mysection{Example}
Here we introduce an example of a Lindblad generator which can be defined on any lattice and generates a renormalisation flow which commutes with translations (and hence does no active rescaling). Given this property, the quantity defined in Eq.~\eqref{betadensity} is non-increasing. 
The Lindblad generator is
\begin{equation}
\label{Pmap}
\mc L(\rho) = \sum_{ij} \left[{ U_{ij} \rho U_{ij}^\dagger - \rho }\right],
\end{equation}
where the sum is over all neighboring pairs of spins on the lattice, and $U_{ij}$ denotes the unitary map swapping sites $i$ and $j$.

The interpretation of this generator is that, after a small ``time'' $\epsilon$, we have a probability $\epsilon$ of confusing any given neighbouring pair of sites, hence losing short scale information. In fact, as we will see, a finite time $t$ amounts to erasing information up to a length scale $\ell \propto \sqrt t$. 
Hence the corresponding generator for the flow in terms of the lengthscale $\ell$ is $\mc L_\ell = \frac{\ell }{\varepsilon^2} \mc L$, where $\varepsilon$ is the lattice spacing.

The fixed points of this generator, i.e.\ states $\sigma$ such that $\mc L(\sigma) = 0$, are precisely the states invariant under permutations of any two sites. According to the quantum de Finetti theorem, this implies that any reduced state on a finite number of sites has the form
\(
\int d\mu(\rho) \,\rho^{\otimes N}
\)
where $\mu$ is a probability measure on mixed states. Hence, due to the absence of active rescaling, the exact fixed points all correspond to {\em mean-field} theories. More interesting critical states must be identified through the asymptotic behaviour of the flow under a passive rescaling of the lattice spacing, i.e., in the continuum limit. 

Let's study the effect of the semigroup on a one-dimensional lattice. It is best interpreted in the Heisenberg picture, where it maps observables of the effective coarse-grained theory to actual physical observables of the microscopic theory. For instance, the action of the generator on an observable $\op a_i$ at site $i$ is
\(
\mc L^\dagger(\op a_i) = \op a_{i+1} -2\op a_{i} + \op a_{i-1}. \\
\)
We recognize a finite-difference approximation of a diffusion equation. Therefore, for large ``time'' $t$, we expect that any {\em local} operator of the form $\op a(f) = \sum_i f_i \op a_i$, were $f_i>0$ and $\sum_i f_i = 1$, will be mapped to
\(
e^{t \mc L^\dagger}(\op a(f))~\simeq~\frac 1 {\sqrt{4 \pi t}}\sum_k e^{-\frac{(k-i)^2}{4t}}  \op a_k.
\)
In particular, we see that the physical resolution of our observables filtered by $e^{t\mc L} = e^{\int_\varepsilon^\ell \mc L_u du}$ is indeed $\ell = \varepsilon \sqrt{t} $. 

\begin{figure}
\vspace{0.4cm}
\includegraphics[width=0.9\columnwidth]{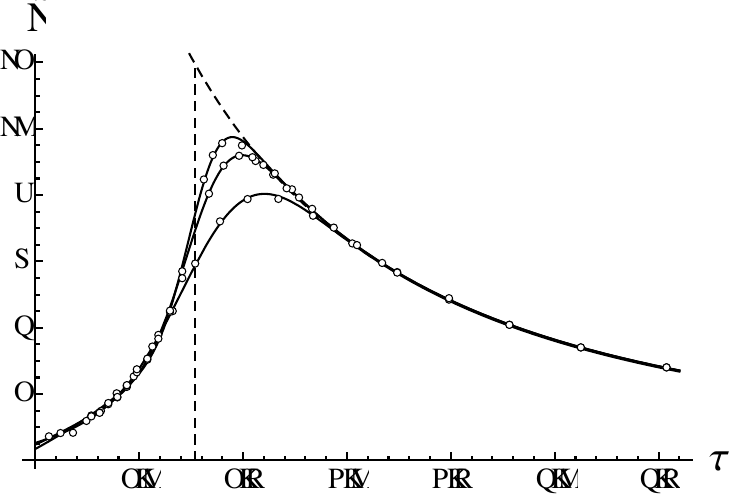} 
\caption{The quantity $\cfun(K)$ defined in Eq.~\eqref{betadensity} (per lattice unit) versus the temperature in unit of $J/k_B$ for the two-dimensional Ising model on periodic square lattices computed using a Monte Carlo simulation. The vertical dashed line represents the exact transition temperature $\tau_c$. The different curves correspond to lattice sizes $10^2$, $20^2$, $40^2$. The fitting function is of the form $a/\tau^{2.1}$ (dashed curve) for $\tau>\tau_c$ and $c - d(\tau_c-\tau)^e$ for $\tau < \tau_c$, convoluted with a Gaussian to account for the finite-size effect. }
\label{mcbeta}
\end{figure}

Figure~\ref{mcbeta} shows the non-increasing quantity defined in Eq.~\eqref{betadensity} for the two-dimensional Ising model on a square-lattice computed using a Monte Carlo simulation. We see that it peaks at the transition temperature, and does not appear to diverge. 

To conclude with this example, let us also show that different fixed points of $\mc L_\ell$ are still distinguishable in terms of the rescaled relative entropy density $D_\ell$, even when the depolarisation generator $\mathcal K_\ell$ introduced above is added to $\mc L_\ell$. 
For instance, consider mean-field states of the form $\rho^{\otimes N}$ and $\sigma^{\otimes N}$, where $N$ is the number of lattice sites, and $\rho$ and $\sigma$ are single-site states. First, observe that $\mc L_\ell$ and $\mc K_\ell$ commute. 
 Since these states are invariant under $\mc L_\ell$, the action of the channel 
 \[
 \mc E_\ell = e^{\int_\varepsilon^\ell (\mc L_u + \mc K_u )du } = e^{\int_\varepsilon^\ell \mc K_u du } e^{\int_\varepsilon^\ell \mc L_u du }
 \]
 on these states is simply given by the depolarizing maps which acts independently on each site. Because the relative entropy is additive on product states, we obtain
\[
D_\ell(\mc E_\ell(\rho^{\otimes \infty})\| \mc E_\ell(\sigma^{\otimes \infty}))  = \frac{\ell^d}{\varepsilon^d} S(\rho_\ell \| \sigma_\ell)
\]
where $\rho_\ell = \tfrac{\varepsilon^d}{\ell^d}\rho  + (1-\tfrac{\varepsilon^d}{\ell^d})\, \tfrac{\one}{\tr \one}$.

Using the joint-concavity of the relative entropy for the upper bound, and the relation $\frac 1 2 (\|\rho-\sigma\|_1)^2 \le S(\rho \| \sigma)$ (See Ref.~\cite{ohya04}) for the lower bound, we obtain
\[
\frac 1 2 ( \| \rho - \sigma \|_1)^2 \;\;\le\;\; D_\ell(\mc E_\ell(\rho^{\otimes \infty})\| \mc E_\ell(\sigma^{\otimes \infty})) \;\;\le\;\;  S(\rho \| \sigma ).
\]
In particular, the lower bound shows that the distance between different mean-field states is asymptotically finite, which shows that this renormalization procedure with ``passive'' rescaling can, at the very least, distinguish between different mean-field phases. 


\mysection{Aknowledgments}
Helpful discussions with Michael Kastoryano and Volkher Scholz are gratefully acknowledged. This work was supported by the ERC grant QFTCMPS and by the cluster of excellence EXC 201 “Quantum Engineering and Space-Time Research”.

\bibliography{renormalization_geometry}

\end{document}